\newcommand{\arxivversion}{true}

\newcommand{\papertitle}{Coherent backscattering of light off
  one-dimensional atomic strings}

\makeatletter \def\namedlabel#1#2{\begingroup
  \def\@currentlabelname{#2}%
  \label{#1}\endgroup
} \makeatother

\documentclass[twocolumn,%
amsmath,amssymb, aps, showpacs, showkeys, floatfix, 10pt, prl,
longbibliography,
nobalancelastpage]{revtex4-1}%

\usepackage[utf8]{inputenc} 

\usepackage[english]{babel} 
\usepackage{csquotes} 

\usepackage{xcolor}%
\usepackage{graphicx}
\usepackage[separate-uncertainty=true]{siunitx} 
\usepackage{bm}

\usepackage{xspace}%
\usepackage{xpatch}

\usepackage{float} %
\usepackage{dcolumn}

\ifdefined\arxivversion\else\ifdefined\paperversion\else
\usepackage{svn-multi} %
\svnid{$Id: atomicmirror.tex 305 2016-07-12 17:41:53Z jappel $} %
\fi\fi

\newcommand{\linkcolor}{magenta}%
\usepackage{hyperref} 
\hypersetup{colorlinks=true, %
  linkcolor=\linkcolor, %
  citecolor=\linkcolor, %
  filecolor=\linkcolor, %
  urlcolor=\linkcolor, %
  pdftitle=\papertitle pdfauthor={H. L. Sørensen, J.-B. Béguin,
    K. W. Kluge, I. Iakoupov, A. S. Sørensen, J. H. Müller,
    E. S. Polzik, J. Appel}, %
  pdfsubject={Quantum optics, Bragg scattering, Tapered optical fiber,
    Nanofiber, Atomic mirror}, %
  pdfkeywords={Quantum optics, Bragg scattering, Tapered optical
    fiber, Nanofiber, Atomic mirror} %
} %

\graphicspath{{images/}}

\makeatletter
\newcommand{\discardpages}[1]{
  \xdef\discard@pages{#1}
  \AtBeginShipout{
    \renewcommand*{\do}[1]{
      \ifnum\pdfstrcmp{\thepage}{##1}=0\relax%
      \AtBeginShipoutDiscard
      \gdef\do####1{}
      \fi%
    }%
    \expandafter\docsvlist\expandafter{\discard@pages}
  }%
} \makeatother

\newcommand{\thesupplementary}[1]{#1}

\newcommand{\nosupplementary}{
  \namedlabel{sec:Detection}{\cite{supp}:SM.A}%
  \namedlabel{sec:ReflectionLifetime}{\cite{supp}:SM.B}%
  \namedlabel{sec:ExperimentalSequence}{\cite{supp}:SM.C}%
  \namedlabel{sec:Transmittance}{\cite{supp}:SM.D}%
  \namedlabel{sec:TransferMatrix}{\cite{supp}:SM.E}%
  \namedlabel{sec:Inhomogeneous}{\cite{supp}:SM.F}%
  \renewcommand{\thesupplementary}[1]{} }

\ifdefined\paperversion%
\ifdefined\withsupplementary%
\def\nosupplementary{}%
\ifdefined\onlysupplementary 
\discardpages{1,2,3,4,5}
\fi \fi%
\else%
\def\nosupplementary{}%
\fi%

\makeatletter%
\def \@labelsection{%
  \@ifundefined{@sectioncntformat}%
  {\@seccntformat}%
  {\@sectioncntformat}{section}%
}%
\def \@labelsubsection{\@labelsection.\thesubsection}%
\def \@labelsubsubsection{\@labelsubsection.\thesubsubsection}%
\xpatchcmd{\@sect@ltx}{\@xsect}{
  \let\@hskip\hskip%
  \def \hskip{\csname hskip \endcsname0.5em plus }%
  \let\@MakeTextUppercase\MakeTextUppercase%
  \def \MakeTextUppercase{}%
  \protected@edef \@currentlabelname{%
    \@hangfrom@section{}{\csname @label#1\endcsname}{#7}%
  }%
  \let\MakeTextUppercase\@MakeTextUppercase%
  \let\hskip\@hskip%
  \@xsect}{}{}
\xpatchcmd{\@ssect@ltx}{\@xsect}{
  \let\@hskip\hskip%
  \def \hskip{\csname hskip \endcsname0.5em plus }%
  \let\@MakeTextUppercase\MakeTextUppercase%
  \def \MakeTextUppercase{}%
  \protected@edef \@currentlabelname{%
    {#7}%
  }%
  \let\MakeTextUppercase\@MakeTextUppercase%
  \let\hskip\@hskip%
  \@xsect}{}{}
\makeatother

\newcommand{\e}{\mathrm{e}}
\newcommand{\ud}{\,\mathrm{d}} 

\newcommand{\avg}[1]{\langle #1\rangle} 
\newcommand{\ket}[1]{\ensuremath{\left|#1\right\rangle}}
\newcommand{\ra}{\rightarrow}

\newcommand{\dtrap}{d_{\text{trap}}} 
\newcommand{\ltrap}{\lambda_{\text{trap}}} 
\newcommand{\ldp}{\lambda_{\text{sp}}} 

\newcommand{\lprobe}{\lambda_{\text{probe}}} 

\newcommand{\fprobe}{\omega_{\text{probe}}} 

\newcommand{\fatom}{\omega_{\text{a}}} 

\newcommand{\omtrap}{\omega_{\text{axial}}} 

\newcommand{\ns}{n_{\text{s}}}

\newcommand{\It}{I_{\text{t}}} \newcommand{\Itn}{I_{\text{bg}}}

\newcommand{\Psat}{P_{\text{sat}}}%
\newcommand{\Na}{N_{\text{a}}}%
\newcommand{\Psp}{P_{\text{sp}}} 
\newcommand{\Pprobe}{P_{\text{probe}}} 

\newcommand{\odperatom}{\alpha_0} 

\newcommand{\sDelta}{\sigma_{\Delta}} 
\newcommand{\ER}{E_{\text{R}}} 
\newcommand{\EL}{E_{\text{L}}} 

\newcommand{\GammaWG}{\Gamma_\text{1D}}

\newcommand{\fref}[2][]{Fig.~\ref{#2}\textcolor{\linkcolor}{#1}} 
\newcommand{\sref}[1]{\nameref{#1}}%

\newcommand{\NBI}{QUANTOP, Niels Bohr Institute, University of
  Copenhagen, Blegdamsvej 17, 2100 Copenhagen, Denmark}

\newcommand{\HLS}[1]{***}

\begin{document}


\newcommand{\correspondingauthors} { \email[Corresponding Authors:
  ]{polzik@nbi.dk} \email{jappel@nbi.dk} \affiliation{\NBI}}

\title{\papertitle}

\author{H. L. S{\o}rensen}%
\author{J.-B. B{\'e}guin}%
\author{K. W. Kluge}%
\author{I. Iakoupov}%
\author{A. S. S{\o}rensen}%
\author{J. H. M{\"u}ller}%
\author{E. S. Polzik} \correspondingauthors%
\author{J. Appel}\correspondingauthors%

\date{\today}

\begin{abstract}
  \ifdefined\svnid {%
    \begin{description}%
    \item[SVN] \footnotesize%
      \textcolor{green}{\svnFullRevision*{\svnrev} by
        \svnFullAuthor*{\svnauthor}, 
        Last changed date: \svndate }%
    \end{description}%
  } \fi%
  
  We present the first experimental realization of coherent Bragg
  scattering off a one-dimensional (1D) system -- two strings of atoms
  strongly coupled to a single photonic mode -- realized by trapping
  atoms in the evanescent field of a tapered optical fiber (TOF),
  which also guides the probe light. We report nearly
  \SI{12}{\percent} power reflection from strings containing only
  about one thousand cesium atoms, an enhancement of two orders of
  magnitude compared to reflection from randomly positioned
  atoms. This result paves the road towards collective strong coupling
  in 1D atom-photon systems. Our approach also allows for a
  straightforward fiber connection between several distant 1D atomic
  crystals.
\end{abstract}


\maketitle

In the search for systems suitable for quantum information technology,
strong light-matter interaction achieved with guided photons has
emerged as one of the favorites. Signatures of efficient coupling of
guided photons have been reported for atoms~\citep{Goban2015} and
quantum dots~\citep{Lodahl:2015fy} coupled to photonic bandgap
waveguides as well as for atoms in a hollow core
fiber~\cite{Bajcsy:2009fx}. Atoms coupled to a
TOF~\citep{LeKien:2004pra,Vetsch2010,Beguin:2014gd} offer strong
coupling of photons to linear strings of about $10^3$ atoms. This has
been predicted to lead to long range
interactions~\citep{Chang:2012co,Chang:2013bh} with prospects to
simulate quantum many-body models~\citep{Douglas:2015hd} and the
ability to generate arbitrary photonic number
states~\citep{Gonzalez-Tudela:2015,Gonzalez-Tudela2016}.

Bragg scattering, well known in crystallography, has become a powerful
tool for artificial atomic structures such as optical
lattices~\cite{Birkl:1995prl,Weidemuller:1995th,Raithel:1997,Deutsch:1995,Slama:2006bf,Schilke:2011fp}. In
free space, 1D optical lattices for atoms can be formed by interfering
two beams and contain between hundreds and thousands of atoms in each
disk-shaped trapping site. Here atoms are much stronger localized in
the axial direction of the standing wave than in the two radial
ones~\citep{Schilke:2011fp,Slama:2006bf}. Such systems have shown
Bragg reflections of \SI{5}{\percent} when probing hundreds of atomic
planes~\citep{Slama:2006bf} and \SI{80}{\percent} when interacting
with $10^7$ atoms located in thousands of
disks~\citep{Schilke:2011fp}. Similarly high reflection efficiencies
have also been observed in periodically structured atomic ensembles in
hot vapor cells~\citep{Bajcsy:2003}.

In this letter, we combine the ideas of waveguiding and Bragg
structuring. We show that with the tight confinement of a guided mode
in a TOF just a thousand atoms are sufficient to create an efficient
1D mirror when arranged as linear strings fulfilling the Bragg
condition.

The basic effect can be understood by considering each atom as a point
scatterer. If the atomic ensemble is spatially unstructured, the
phases of backscattered fields from individual atoms are
uncorrelated. Waves emitted backwards thus add up incoherently and,
for an optically thin sample, the total intensity is proportional to
the number of scatterers. If, on the other hand, the atoms are
arranged under the Bragg condition, spaced distances $d = q
\lprobe^\text{TOF}/2$ apart, where $q$ is an integer and
$\lprobe^\text{TOF}$ is the wavelength of the probe light within the
TOF \footnote{throughout the paper $\lambda$ with (without) the
  superscript ``TOF'' refers to the TOF (free-space) wavelength},
there is a fixed phase relation. This enhances the intensity by
constructive interference and, in the limit of perfect arrangement,
the reflected power scales quadratically with the number of scatterers
for an optically thin medium.


In the experiment, we trap cesium atoms in a dual-color TOF-based
lattice
trap~\citep{LeKien:2004pra,Vetsch2010,Goban:2012ct,Beguin:2014gd}
(details in \sref{sec:ExperimentalSequence}), which holds the atoms as
two 1D strings \SI{200}{\nano \meter} above the surface on opposite
sides of the fiber (see \fref[a,b]{fig:setup}).
\begin{figure*}[htp] 
  \includegraphics[width=0.95\textwidth]{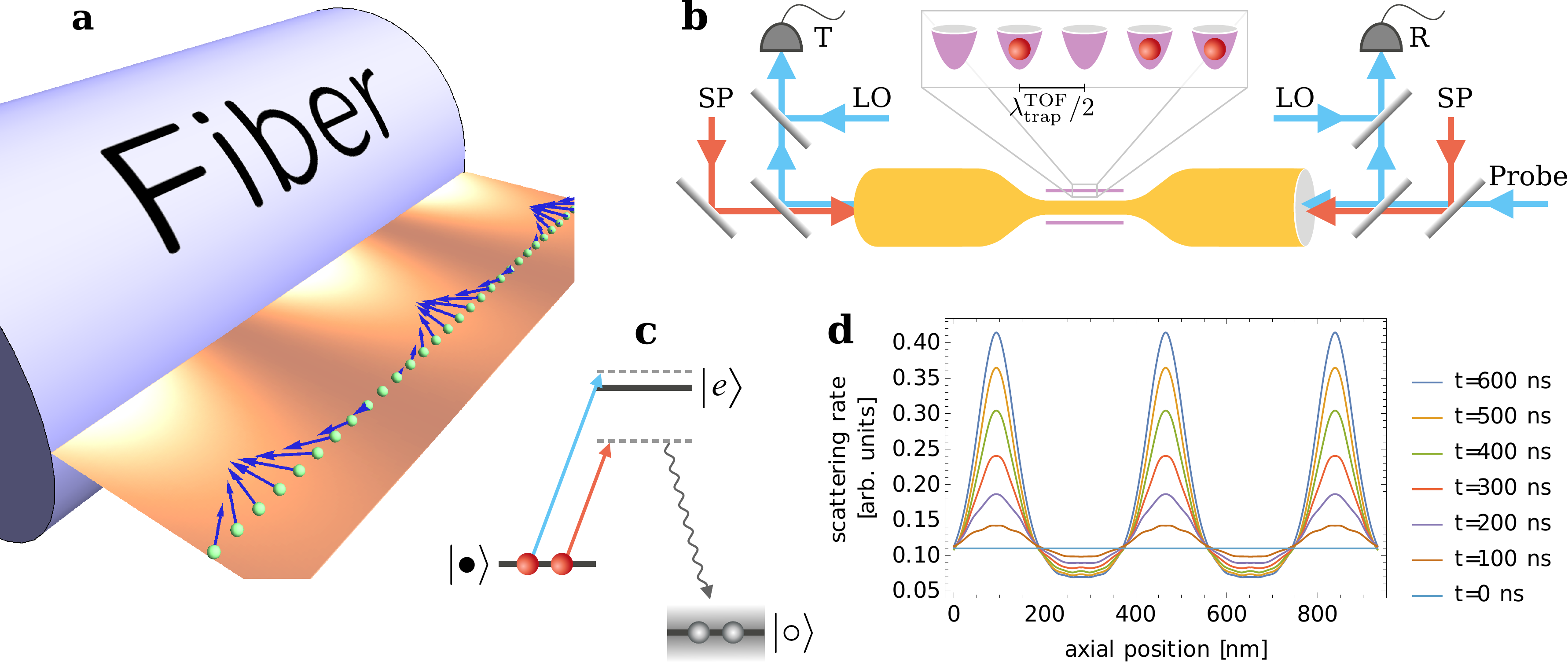}
  \caption{\label{fig:setup} Experimental procedure and
    setup. \textbf{a}, The standing wave structuring pulse field
    (orange plane) accelerates atoms (green balls) both axially and
    radially towards the intensity maxima (bright spots). Since the
    trapping and structuring fields have incommensurate wavelengths,
    we depict with blue arrows the positions of atoms starting at the
    radial equilibrium location and various axial positions after a
    ballistic flight of $\SI{600}{\nano\second}$ following the
    structuring pulse. \textbf{b}, Experimental setup. In the dipole
    trap, the atoms arrange themselves into two 1D strings (purple
    lines) along the TOF (yellow). Each trap site (purple wells in the
    zoom box) is either empty or contains a single atom prepared in
    the $\ket{\bullet}$ state. A standing wave pulse (SP, red) is used
    to imprint a Bragg grating onto the atoms. A probe field (blue) is
    sent onto the atomic strings; the reflected and transmitted light
    is mixed with corresponding local oscillator (LO) beams and
    measured by two photo-detectors R and T (heterodyne
    detection). \textbf{c}, Level diagram. The probe beam (blue) is
    tuned close to resonance with the $\ket{\bullet}\ra\ket{e} \equiv
    (6^2P_{3/2},F=5)$ transition. The structuring pulse (red)
    optically exerts both a radial and axial dipole force and pumps
    the atoms out of \ket{\bullet}. \textbf{d}, Evolution of the axial
    modulation of the scattering rate, calculated for an atomic
    ensemble located at the trap potential minimum (green balls in
    \textbf{a}) with an initial thermal velocity distribution
    corresponding to $T=\SI{42}{\micro\kelvin}$}.
\end{figure*}
The distance between neighboring trap sites, which contain at most a
single atom due to atomic collisions during the loading process
\cite{Schlosser:2002}, is set by the wavelength of the attracting
standing wave trapping laser $\dtrap=\ltrap^\text{TOF}/2$. To avoid
trap-induced scattering, $\ltrap = \SI{1057}{\nano\meter}$ is
far-detuned from atomic resonance, whereas the probe light, at a
vacuum wavelength $\lprobe=\SI{852}{\nano\meter}$, is near atomic
resonance to ensure appreciable backscattering. The corresponding TOF
wavelengths due to the fiber modal dispersion are $\ltrap^\text{TOF} =
\SI{987}{\nano\meter}$ and $\lprobe^\text{TOF} =
\SI{745}{\nano\meter}$. With a sample length of
$\sim\SI{1}{\milli\meter}$ the incommensurate ratio
$\ltrap^\text{TOF}/\lprobe^\text{TOF}$ allows us to treat the atomic
strings as originally completely unstructured with respect to the
Bragg resonance, leading to a weak reflection as observed
in~\cite{Reitz:2014cxa}.

Initially, all atoms are prepared in the same electronic ground state
$\ket{\bullet} \equiv (6^2S_{1/2},F=4)$. To turn the atomic strings
into an effective Bragg mirror, the ensemble is structured with a
short (\SI{250}{\nano\second}) standing wave light pulse while the
two-color TOF trap remains active. The frequency of the structuring
light is detuned by $\SI{-175}{MHz}$ from the
$\ket{\bullet}\rightarrow (6^2P_{3/2},F=3)$ transition to allow for
propagation through the ensemble, which is opaque for resonant light
(\fref[a]{fig:setup} and~\fref[c]{fig:setup}).

The quasi-linearly vertically polarized structuring and probe fields
possess vanishing longitudinal field components at the trapping sites,
which ensures a high contrast of the structuring intensity grating
experienced by the atoms. This choice of probe polarization leads to
an equal probability of scattered fields to couple into either the
forward or backward TOF-guided mode, whereas for spin-preserving
scatterers, the use of orthogonally polarized probe light would lead
to predominant forward-scattering~\cite{LeKien:2014hy}.

As the standing wave light pulse is sent through the TOF, it affects
all atoms except the ones localized at its nodes. The coupling of the
trapped atoms to the running wave probe field mode is changed by the
structuring pulse through modulation of the atomic electronic and
motional state (see \fref[d]{fig:setup}). Various physical effects,
such as hyperfine pumping, Zeeman-level pumping, axial and radial
acceleration due to dipole forces contribute to this modulation.  We
emphasize that the two-color TOF dipole trap is active during the
entire structuring and Bragg scattering measurement sequence.

With the detuning and duration of the structuring light pulses chosen
in this work, the effect of the dipole forces is dominant. For a
structuring pulse tuned below the atomic resonance, atoms located
distant to intensity nodes receive momentum radially towards the fiber
and axially towards the closest antinode (see
\fref[a]{fig:setup}). After a short time, this imprinted velocity
grating transforms into a density- and coupling-modulation with maxima
separated by multiples of half-wavelengths of the structuring pulse,
satisfying the Bragg condition in close analogy to a photo-refractive
medium with a refractive index grating \footnote{The small relative
  wavelength difference between probe and structuring light
  $\delta\lambda/\lambda = 5 \cdot 10^{-7}$ leads only to a negligible
  phase mismatch over the sample length ($\SI{1}{\milli\meter} \sim
  10^3 \lambda)$}.

We measure the reflected and transmitted fields simultaneously by
optical heterodyne detection (\fref[b]{fig:setup}) in an (adjustable)
bandwidth of \SI{9.2}{\mega\hertz} (details in \sref{sec:Detection})
to avoid being affected by other stray light sources (such as the trap
fields). As the effective mode area of the probe at the position of
the atoms is small~\cite{LeKien:2006bh}, we use an extremely weak
probe field with a typical power of
$P_{\text{incident}}=\SI{150}{\pico\watt}$ to avoid saturation.

\begin{figure}[tb!]
  \includegraphics[width=\columnwidth]{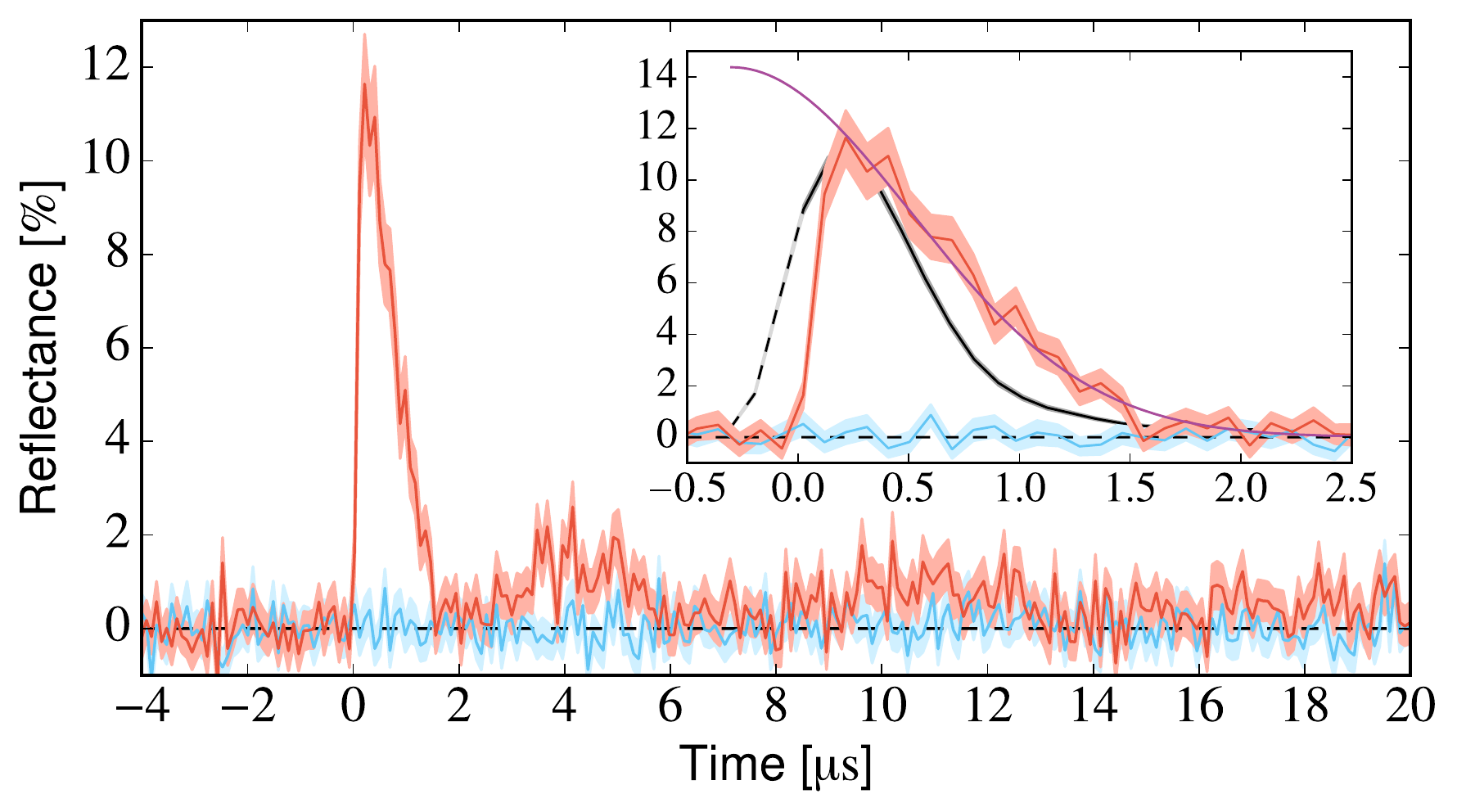}
  \caption{\label{fig:timetraces} Reflectance off the atomic strings
    within a \SI{2.8}{\mega\hertz} detection bandwidth
    (\SI{3}{\decibel}). The probe is tuned \SI{8}{\mega\hertz} above
    atomic resonance and turned on at $t=0$ with an independently
    measured rise time of \SI{80}{\nano\second}. Blue curve:
    unstructured atomic ensemble, average of 250 experiments. Red
    curve: structured ensemble, average of 200 experiments. The shaded
    regions signify the one-sigma uncertainty interval with
    contributions from statistical averaging and a \SI{5}{\percent}
    input probe power fluctuation during the measurement. Inset: zoom
    on the first reflection peak with a fit to a Gaussian decay
    (purple) and the theoretical prediction (black) with uncertainty
    band given by the statistical averaging.}
\end{figure}

In \fref{fig:timetraces}, we present a typical example of the temporal
dynamics of the reflectance (defined as the ratio of reflected to
incident power $P_{\text{reflected}}/P_{\text{incident}}$), for the
structuring pulse present and absent. The curves are averaged over
multiple consecutive experimental runs to overcome the shot noise due
to the low photon flux in single experiments. With the incident probe
power chosen well below saturation (see above), for every 60 input
photons, on average only 4 photons impinge onto the reflection
detector within the \SI{96}{\nano\second} sample time at the peak of
the curve.

The probe is turned on at $t=0$ shortly after the structuring pulse
has been turned off at $t_0=-\SI{0.3}{\micro\second}$. The peak power
reflectance of $\SI{11.6 \pm 1.1}{\percent}$ from the structured
atoms, shown in \fref{fig:timetraces}, is two orders of magnitude
stronger than the average reflectance of $\SI{0.10 \pm
  0.01}{\percent}$ observed from 1900 realizations of an unstructured
ensemble.

As shown in the inset of~\fref{fig:timetraces}, the decay after the
maximum is well fitted by a Gaussian function \mbox{$f(t) =
  f(t_0)\exp[-(t-t_0)^2/2\tau^2]$} with a characteristic decay time
\mbox{$\tau =\SI{0.82 \pm 0.02}{\micro\second}$}. We use the value of
the fit function at $t = \SI{0.2}{\micro\second}$ to extract robust
values for the peak reflectance from the data. Physically, the
reflectance depends on the degree of localization of atoms around the
proper axial locations to fulfill the Bragg condition. The spatial
ordering first builds up and then decays as atoms move past the focus
points at the antinodes of the structuring wave pulse. In addition,
any thermal random initial velocity on top of the deterministic
imprinted velocity grating limits the quality of localization and
accelerates the decay of ordering. After structuring, the atoms are
left in a highly nonthermal motional state within the trap potential
wells.  We have verified that the weak probe laser itself does not
influence the decay time of the reflectance signal
(see~\ref{sec:ReflectionLifetime}).

Besides the significant reflection off the structured atoms following
the probe onset, two much smaller peaks are visible after roughly
\SI{4}{\micro\second} and \SI{11}{\micro\second} in
\fref{fig:timetraces}. Reaching reflectances of \SI{2.6 \pm
  0.5}{\percent} and \SI{1.9 \pm 0.6}{\percent} respectively, these
signals are clearly distinguishable from a reflection off an
unstructured ensemble. We attribute them to partial revivals of
localization at the Bragg condition for the atomic wave packets
sloshing and breathing in the incommensurate lattice of anharmonic
trap wells.

For theoretical modeling, in separate measurements we determine the
number of trapped atoms $\Na$ for calibration purposes by counting the
number of scattered photons required to optically pump all the atoms
from $\ket{\bullet}$ to $(6^2S_{1/2},F=3)$~\cite{Beguin:2014gd}. From
the temporal dynamics of the pumping process, we can also extract the
single-atom optical depth $\odperatom$, albeit with a large relative
uncertainty of about \SI{25}{\percent}, as such a measurement is
affected by an unequal population distribution within the Zeeman
sub-levels of \ket{\bullet} and broadening of the optical transitions
due to the trapping fields. We obtain
$\odperatom=\SI{0.51\pm0.13}{\percent}$, corresponding to an
on-resonant saturation power of $\Psat=
\Gamma\hbar\omega/(2\odperatom) = \SI{750}{pW}$ (i.e. $\sim 100$
photons per excited state lifetime) for the $\ket{\bullet}\ra
\ket{e},\Delta m_F=0$ transition.
The theoretical prediction for the overall reflectance is modeled by
employing the 1D transfer matrix
formalism~\citep{Deutsch:1995,Chang:2012co}: An initially random
string of atoms is structured by spatially modulated dipole forces and
hyperfine pumping. Its reflectance for probe light at any given time
is evaluated by multiplying matrices describing the scattering off
individual atoms alternating with matrices for free propagation
(see~\sref{sec:TransferMatrix} for details). Inhomogeneous broadening
of the probe transition by the strong trap light fields is taken into
account by assigning a random Gaussian distributed detuning to each
atom with a standard deviation of $\sDelta = 0.41 \Gamma$, where
$\Gamma = 2\pi\times\SI{5.23}{\mega\hertz}$ denotes the natural
linewidth of the atomic transition (see \sref{sec:Inhomogeneous} for
details). The coupling of atoms to TOF-guided probe light at their
radial equilibrium position is set to the experimentally determined
value of $\avg{\odperatom}=\SI{0.51}{\percent}$ for the on-resonant
optical depth per atom and scaled with the radial mode
shape~\cite{Kien:2004hh}. The initial radial position and velocity are
randomized according to a thermal distribution with $\avg{T}
=\SI{42}{\micro\kelvin}$, about one fifth of the
\SI{200}{\micro\kelvin} trap well depth. Radial motion due to the
dipole force takes place in a harmonic oscillator potential with the
independently measured radial trap frequency of
$\SI{85}{\kilo\hertz}$, truncated at $\SI{10}{\nano\meter}$ from the
TOF surface where atoms are irretrievably lost. For simplicity, dipole
force induced axial motion is treated ballistically, which is
justified since the observed dephasing times are much shorter than an
axial oscillation cycle in the true trap potential (\mbox{$\omtrap
  \simeq \SI{100}{\kilo\hertz}$}). In accordance with the experimental
procedure, we average the simulation results over typically $100$
random realizations.

In the inset in \fref{fig:timetraces} the theoretical prediction for
the reflection is shown to be in quantitative agreement with the peak
reflectance. The faster temporal decay can be assigned to the
simplified model not taking the true trap potential into account.

The observed critical dependence of reflectance on the degree of
localization also strongly influences the conditions for achieving
optimal reflectance. For a stronger structuring pulse the atomic wave
packets become focused and localized faster, hence outrunning the
dephasing by thermal motion. For a too strong structuring pulse,
however, the duration of high reflectivity decreases since a
significant fraction of atoms climbs over the repulsive barrier of the
trap and crashes into the hot fiber surface rapidly. This results in
an optimal structuring pulse power due to the trade-off between a high
degree of localization and a high number of remaining scatterers.

To visualize this trade-off, experimental data for the reflectance as
a function of the structuring power $\Psp$ is shown in
\fref[a]{fig:R_vs_detuning_and_fraction}. Immediately before taking
the data series, $\Na$ was measured to be 1440. To confirm a stable
initial atom number, $\Na$ was measured again right after taking the
data series to 1140. From this we assign $\Na=1290\pm150$. Both the
experiment and the model calculations, using the same input parameters
for the simulation as above, show an optimum sample reflectance. The
experimentally measured slopes on either side of the optimum are well
captured by the model.

\begin{figure}[tb]
  \includegraphics[width=\columnwidth]{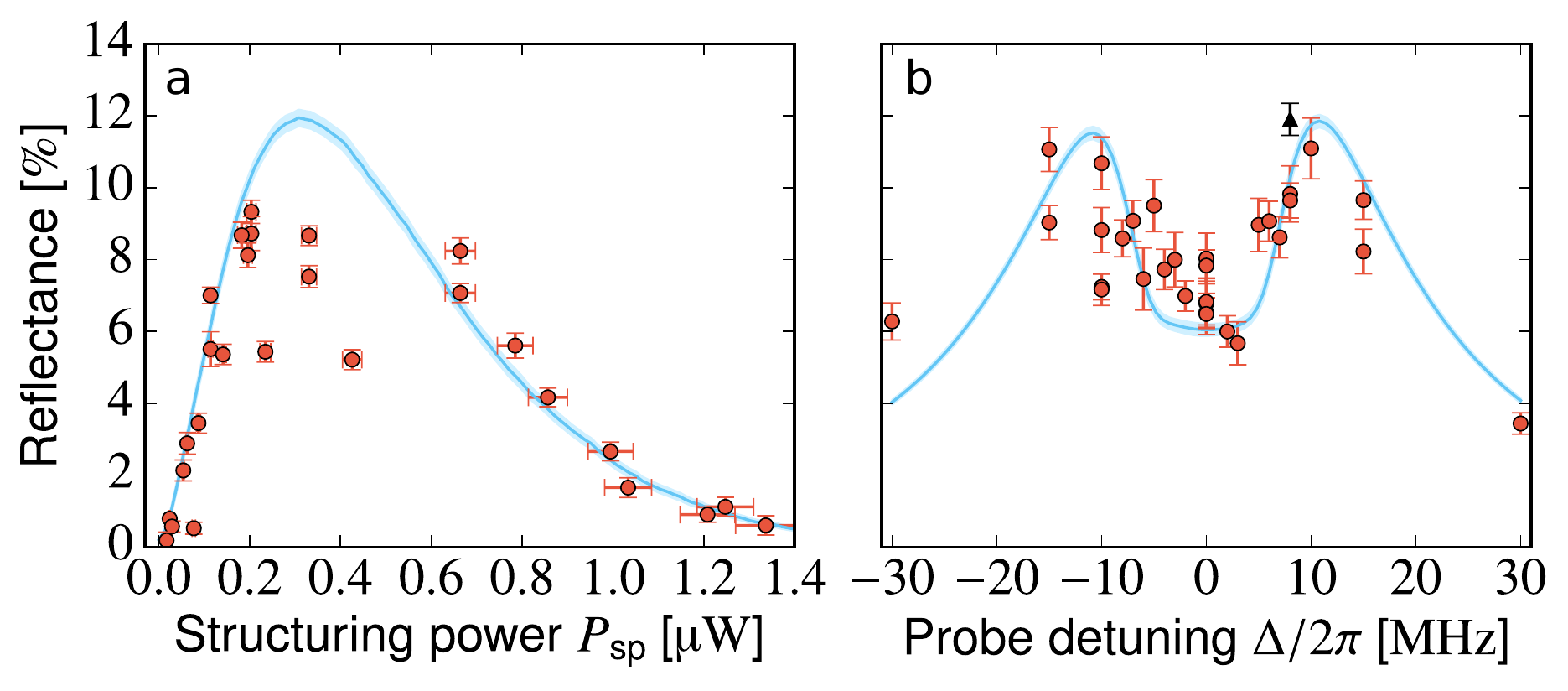}
  \caption{\label{fig:R_vs_detuning_and_fraction} Experimental data.
    Each point represents the reflectance at
    $t=\SI{0.2}{\micro\second}$ extracted from a Gaussian decay fit to
    100--250 consecutive experimental runs with vertical error bars
    from the uncertainty on the fit parameters. Blue curves show the
    theoretical predictions with the shaded regions signifying the
    one-sigma uncertainty interval from the statistical averaging.
    \textbf{a}, Reflectance as a function of the power of the
    structuring pulse $\Psp$ for a probe detuning of
    $\Delta=2\pi\times\SI{8}{\mega\hertz}$. For this data
    $\avg{\tau}=\SI{0.89}{\micro\second}$. Horizontal error bars
    corresponds to a \SI{5}{\percent} power fluctuation. \textbf{b},
    Variation of the reflectance with the probe detuning. The Bragg
    grating was created using a structuring power of
    \SI{0.2}{\micro\watt}.  The black triangular point corresponds to
    the red data trace in \fref{fig:timetraces}. For this data
    $\avg{\tau}=\SI{0.86}{\micro\second}$.}
\end{figure}

The influence of incomplete atomic localization is also evident when
interrogating the atomic strings at varying detunings,
$\Delta=\fprobe-\fatom$, of the probe laser from the atomic
transition. If atoms were perfectly positioned, such as to fulfill the
Bragg condition, the incident probe and backscattered fields would
form a standing wave with nodes coinciding with the atom
locations~\citep{Slama:2006bf}. This reduces the absorption and allows
the probe to propagate further into the sample, thereby increasing the
number of scatterers contributing to the reflected field. In this
case, the reflection spectrum is expected to be a Lorentzian with a
maximum reflectance on resonance where the light-atom coupling is
strongest~\cite{LeKien:2014eb}. When delocalized atoms are present,
this description breaks down since diffuse scattering becomes
prominent and increases the absorption. The penetration length of the
probe thus becomes drastically reduced and the number of scatterers is
effectively lowered. This can be circumvented by tuning the probe
frequency slightly off-resonance and consequently decreasing the
light-atom coupling~\cite{Birkl:1995prl,Slama:2006bf}. The detuned
probe field then propagates further into the sample allowing for
interaction with more atoms, which results in an increase in the
coherent backscattering. In this regime, a reflection spectrum is
expected to have a dip on resonance and maxima at probe detunings
where the scattering probability is optimally balanced for high
penetration depth. Experimental data confirming this is presented
in~\fref[b]{fig:R_vs_detuning_and_fraction}: The coherent Bragg
reflection is the highest for detunings between \numrange{1.5}{3} line
widths from resonance. Using the same input parameters as above for
the simulation we find satisfactory quantitative agreement between the
data and the model.

In summary, we have shown that a truly one-dimensional dilute Bragg
grating containing in total around a thousand atoms reflects more
light than a solid slab of glass, reaching reflectances of more than
\SI{10}{\percent}. The TOF trap platform offers exciting opportunities
for future developments. The presented concept of creating a
switchable Bragg-reflector by modulating the probe light coupling with
standing-wave structuring light fields is versatile: instead of
modifying the motional degrees of freedom of the atoms, optical
pumping into other electronic states or coherent population transfer
using a Raman process open paths to increase the lifetime of the
reflection peak. Motional dephasing can be suppressed with colder
atomic samples. The simulations predict that by applying the same
structuring procedure at a temperature of $T =\SI{2}{\micro\kelvin}$
the reflectance can be doubled. Trapping the atoms permanently closer
to the surface will increase the coupling and using a longer tapered
fiber section will increase the overall atom number thus improving the
reflectance. Adding an optical cavity with a modest finesse integrated
with the fiber~\cite{Kato:2015is} will boost the single atom
reflectivity to near unity. Such a cavity will reduce the present
saturation photon number of about a hundred to the single photon
level, thus allowing for strong single photon-light coupling in this
1D system. For these high coupling strengths, exploration and
exploitation of atomic self-organization
mechanisms~\citep{Chang:2013bh} becomes an attractive playground for
experiments. With these improvements the creation of resonantly
structured atomic ensembles in the vicinity of photonic waveguides
allows for new avenues in quantum state
preparation~\citep{Gonzalez-Tudela:2015} as well as for the formation
of atomic cavities~\citep{Chang:2012co}.

\begin{acknowledgments}
  This work was supported by the ERC grants INTERFACE (grant
  no. ERC-2011-ADG 20110209) and QIOS (grant no. 306576), and the EU
  project SIQS (grant no. 600645). The authors would like to thank
  E.~M. Bookjans for help in setting up the experiment.
\end{acknowledgments}

\medskip \emph{Note added} – During submission of our manuscript, we
became aware of a new related study~\citep{Corzo2016}.

\ifdefined\arxivversion\else\ifdefined\paperversion\else%
\InputIfFileExists{texcount.aux}{}\relax%
\fi\fi

\bibliographystyle{apsrev4-1} 
\bibliography{atomicmirror}


\nosupplementary

\thesupplementary{ \newpage \clearpage

\ifdefined\arxivversion \else
  \setcounter{page}{1} 
  \renewcommand{\thepage}{S\arabic{page}} 
  \fi

  \appendix

  \setcounter{equation}{0} 
  \renewcommand{\theequation}{\Alph{section}\arabic{equation}} 
  \renewcommand{\theHequation}{\theequation} 

  \setcounter{figure}{0} 
  \renewcommand{\thefigure}{S\arabic{figure}} 
  \renewcommand{\theHfigure}{\thefigure} 

  \renewcommand{\thesection}{SM.\Alph{section}}
  \setcounter{secnumdepth}{1}

  \section*{Supplementary Material}
  \renewcommand{\appendixname}{}
  \label{sec:Supplemental}
  In the following, we provide additional information on experimental
  details and the numerical implementation of the theoretical model.
  In \sref{sec:Detection}, we provide technical details on our
  detecion method.  In \sref{sec:ReflectionLifetime} we discuss
  influence by the probe on the temporal decay of the reflectance. The
  full experimental sequence is presented in
  \sref{sec:ExperimentalSequence} followed by \sref{sec:Transmittance}
  where additional experimental data for the sample transmittance is
  presented and discussed. The numerical procedure to obtain the
  theoretical predictions for the reflectance is specified in
  \sref{sec:TransferMatrix}, and finally, we conclude with
  \sref{sec:Inhomogeneous}, in which we show how the amount of
  inhomogeneous broadening is estimated.

  \section[]{Detection}%
  \label{sec:Detection}
  In the experiment, we measure the reflected and transmitted fields
  simultaneously by optical heterodyne detection
  (\fref[b]{fig:setup}): Before entering the TOF, the probe is split
  from a strong ($\approx \SI{700}{\micro\watt}$ for the reflection
  detector) local oscillator (LO) reference beam and sent through an
  acousto-optical modulator to shift it in frequency by
  $\Omega=2\pi\times\SI{-62.5}{\mega\hertz}$. The transmitted and
  reflected fields are each superimposed with their corresponding LO
  using 90:10 beam splitters and detected with low-noise
  AC-photo-detectors. The recorded quadrature signals of the beat-note
  are digitally demodulated, low-pass filtered and converted into a
  photon flux, using independently both a calibrated power reference
  and the LO shot-noise level.

  With a detection bandwidth of
  $B_{\text{3dB}}=0.89/t_\text{sample}=\SI{9.2}{MHz}$, we determine
  the number of signal photons hitting the detector during the sample
  time $t_\text{sample}=\SI{96}{\nano\second}$ by
  \begin{equation}
    \ns = \frac{1}{\eta \mathcal{V}^2} \frac{
      \delta i_{\text{signal}}^2 - \delta i_\text{sn+en}^2}{
      \delta i_\text{sn+en}^2 - \delta i^2_{\text{en}}},
  \end{equation}
  where $\eta=0.89$ is the quantum efficiency of the detector,
  $\mathcal{V}=0.93$ is the measured interferometric visibility.  The
  photocurrent signal $s(t)$ is mixed down to base band and averaged
  over a time $t_\text{sample}$ to obtain $\delta
  i_{\text{signal}}^2(t) = \left| \int_{t}^{t+t_\text{sample}}
    \e^{\operatorname{i} \Omega t'} s(t')
    \frac{\operatorname{d}\!t'}{t_\text{sample}} \right|^2 $.  Its
  value with blocked probe input is $\delta i_\text{sn+en}^2$, and
  $\delta i_{\text{en}}^2$ denotes the electronic noise of the
  detector as measured with both the probe- and the LO-beam blocked.

  Including the amount of optical losses in the path from the atomic
  strings to the reflection detector of \SI{54}{\percent}, we arrive
  at the number of photons reflected off the atoms.

  \section[]{Reflection lifetime}
  \label{sec:ReflectionLifetime}
  In \fref{fig:tau} we show the histogram of the fitted characteristic
  time constant $\tau$ of the reflectance decay according to the
  Gaussian model function given in the main text.
  \begin{figure}[htb]
    \includegraphics[width=\columnwidth]{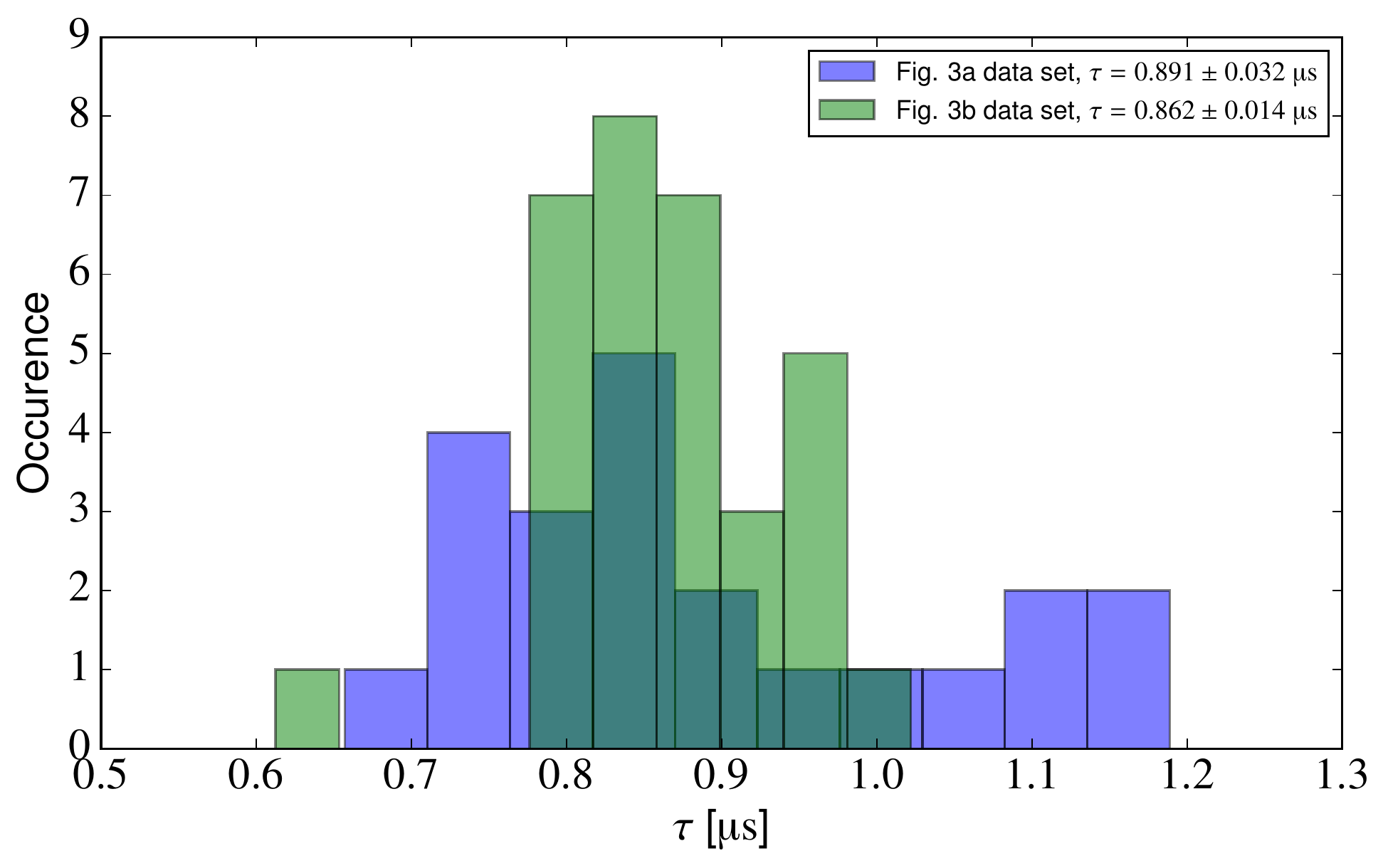}
    \caption{Histogram of $\tau$ for the two data sets used in
      \fref[a,b]{fig:R_vs_detuning_and_fraction}. In the legend the
      mean value and the standard deviation of the mean of $\tau$ is
      given. \label{fig:tau} }
  \end{figure}

  To verify experimentally that the decay is not probe induced, we
  have performed a series of measurements with increasing delay
  between the end of the structuring pulse and the start of probe
  pulse as shown in \fref{fig:probedelay}.
  \begin{figure}[htbp]
    \includegraphics[width=\columnwidth]{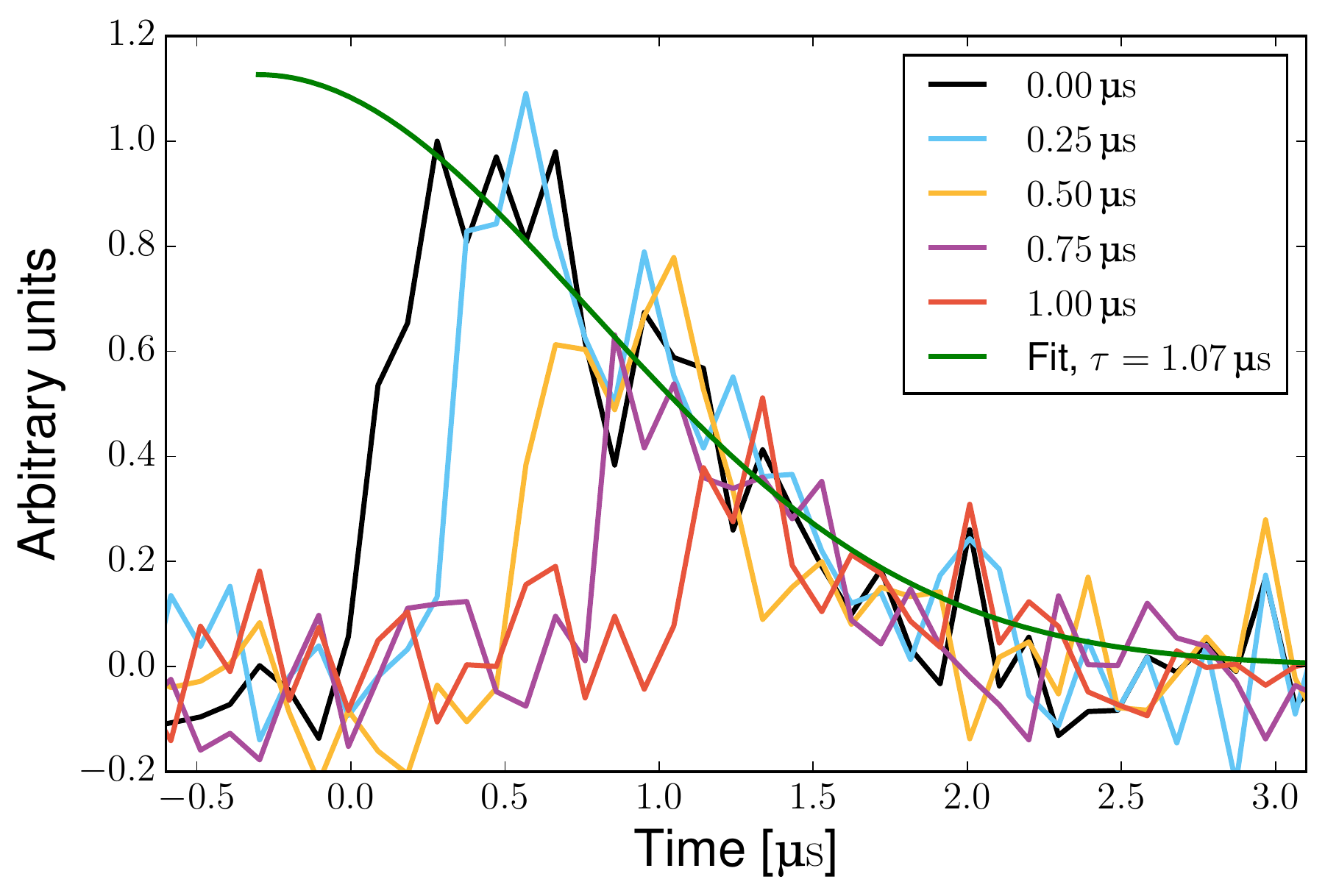}
    \caption{Result of the reflection measurements with delayed probe
      pulses. Error bars omitted for visual clarity. Each curve is an
      average over 100 consecutive experimental runs obtained for a
      $\SI{150}{pW}$ on-resonant probe.\label{fig:probedelay} }
  \end{figure}
  If the reflectance lifetime were dominated by a destructive
  influence from the probe, we would expect to reach the same amount
  of reflection for the various time delays. This is clearly not the
  case. In \fref{fig:probedelay} we see that the decaying tails of the
  probe signals all follow the same global curve described by a single
  Gaussian decay fit to the black curve with probe onset at $t=0$.  We
  thus conclude that the probe itself does not have a significant
  influence on the dephasing of the imprinted Bragg grating. We obtain
  the same results for greater probe powers and off-resonant probe
  light as well (not shown).

  \section[]{Experimental sequence}
  \label{sec:ExperimentalSequence}
  In each experiment, we first load cesium atoms from a background
  vapor into a standard 6-beam magneto-optical trap (MOT). This is
  followed by sub-Doppler cooling during which the atoms are loaded
  into a dual-color TOF-based
  trap~\citep{LeKien:2004pra,Vetsch2010,Goban:2012ct,Beguin:2014gd};
  using two quasi-linearly horizontally polarized counter-propagating
  red-detuned fields with wavelength
  $\lambda_{\text{trap}}=\SI{1057}{\nano\meter}$ and power
  $P_{\text{trap}}\approx2\times\SI{1.3}{\milli\watt}$ together with a
  quasi-linearly vertically polarized blue-detuned field (with
  wavelength $\lambda_{\text{blue}}=\SI{780}{\nano\meter}$ and power
  $P_{\text{blue}}=\SI{14}{\milli\watt}$) optical dipole traps are
  formed about $\SI{200}{\nano\meter}$ above the TOF surface, which
  hold the atoms as two 1D strings on both sides of the fiber (see
  \fref[a]{fig:setup}). After loading the atoms into this trap, the
  MOT beams are shut off and a waiting time of \SI{11}{\milli\second}
  is imposed to ensure that only trapped atoms are probed.

  In \fref{fig:exp_sequence} the timing for the different experimental
  steps for preparing and probing the single atomic Bragg grating is
  shown. The initial MOT loading and subsequent sub-Doppler cooling
  for transferring the atoms into the TOF trap takes $\sim
  \SI{2}{\second}$.  Prior to creating the Bragg grating an initial
  reference measurement for the fully transmitted field with beat note
  intensity $I_0$ is established while having all the atoms in the
  transparent state $\ket{\circ}$ (from $t=\SI{-10}{\micro\second}$ to
  $t=\SI{-6}{\micro\second}$) (see also \fref{fig:trm}).
  \begin{figure}[htbp]
    \includegraphics[width=\columnwidth]{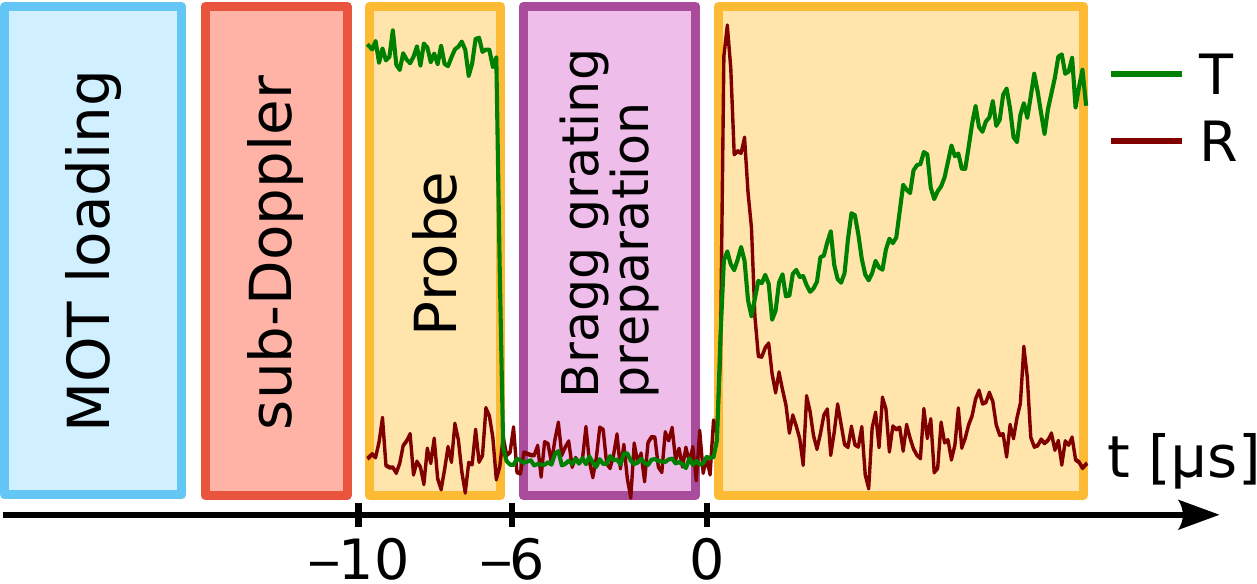}
    \caption{\label{fig:exp_sequence} Time sequence of a single
      experimental run. Green (red) curve is an illustration of the
      transmitted (reflected) power (not to scale).}
  \end{figure}
  Immediately after, the atoms are optically pumped from \ket{\circ}
  to \ket{\bullet} in a duration of $\SI{3}{\micro\second}$. This is
  followed by the structuring pulse creating the transient gradient of
  the light atom coupling. At $t=0$ the probe is turned back
  on. Measurements of both the reflected and transmitted fields are
  continuously carried out from $t=\SI{-10}{\micro\second}$ to
  $t=\SI{40}{\micro\second}$.

  \section[]{Transmittance}
  \label{sec:Transmittance}
  From the detected beat note intensity $\It$ of the transmitted
  light, the detuning dependent optical depth $\alpha$ can be
  extracted via Lambert-Beer's law; $\mathcal{T}=\e^{-\alpha}$, where
  \begin{equation}
    \mathcal{T} = \frac{\It - \Itn}{I_0 - \Itn }
  \end{equation}
  is the transmittance, see \fref{fig:trm}, and $\Itn$ is the
  detection background when the probe is off from
  $t=\SI{-6}{\micro\second}$ to $t=0$ as shown in
  \fref{fig:exp_sequence}.
  \begin{figure}[htbp]
    \includegraphics[width=\columnwidth]{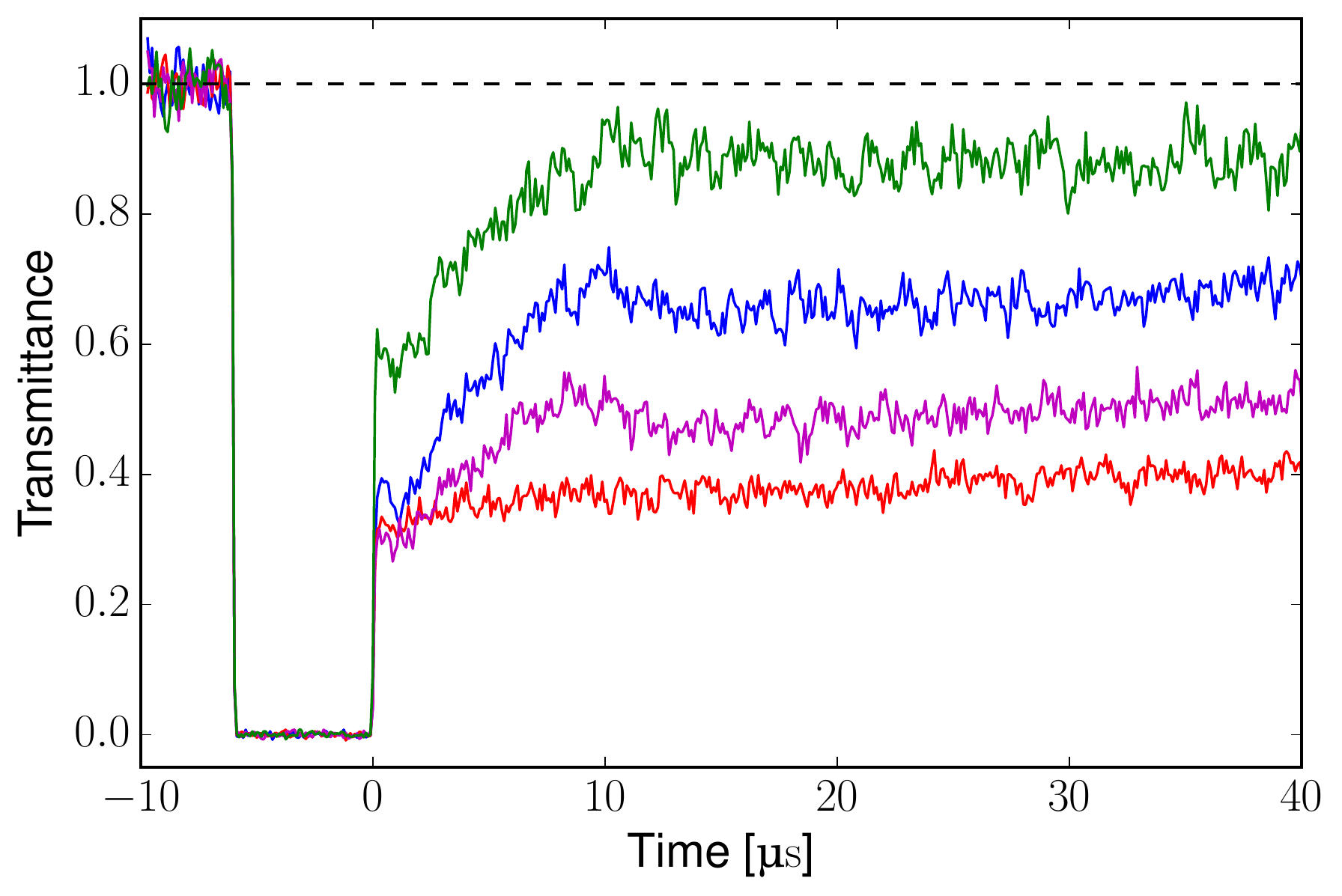}
    \caption{\label{fig:trm} Transmittance. The traces are obtained
      for $\Pprobe=\SI{140}{\pico\watt}$, $\Delta=\SI{8}{MHz}$,
      $\Na\approx 1300$, and four different powers of the structuring
      pulse $\Psp$: Green: $\Psp= 2\times\SI{1.0}{\micro\watt}$, Blue:
      $\Psp=2\times\SI{0.2}{\micro\watt}$, Purple:
      $\Psp=2\times\SI{0.09}{\micro\watt}$, Red: no structuring.}
    \includegraphics[width=\columnwidth]{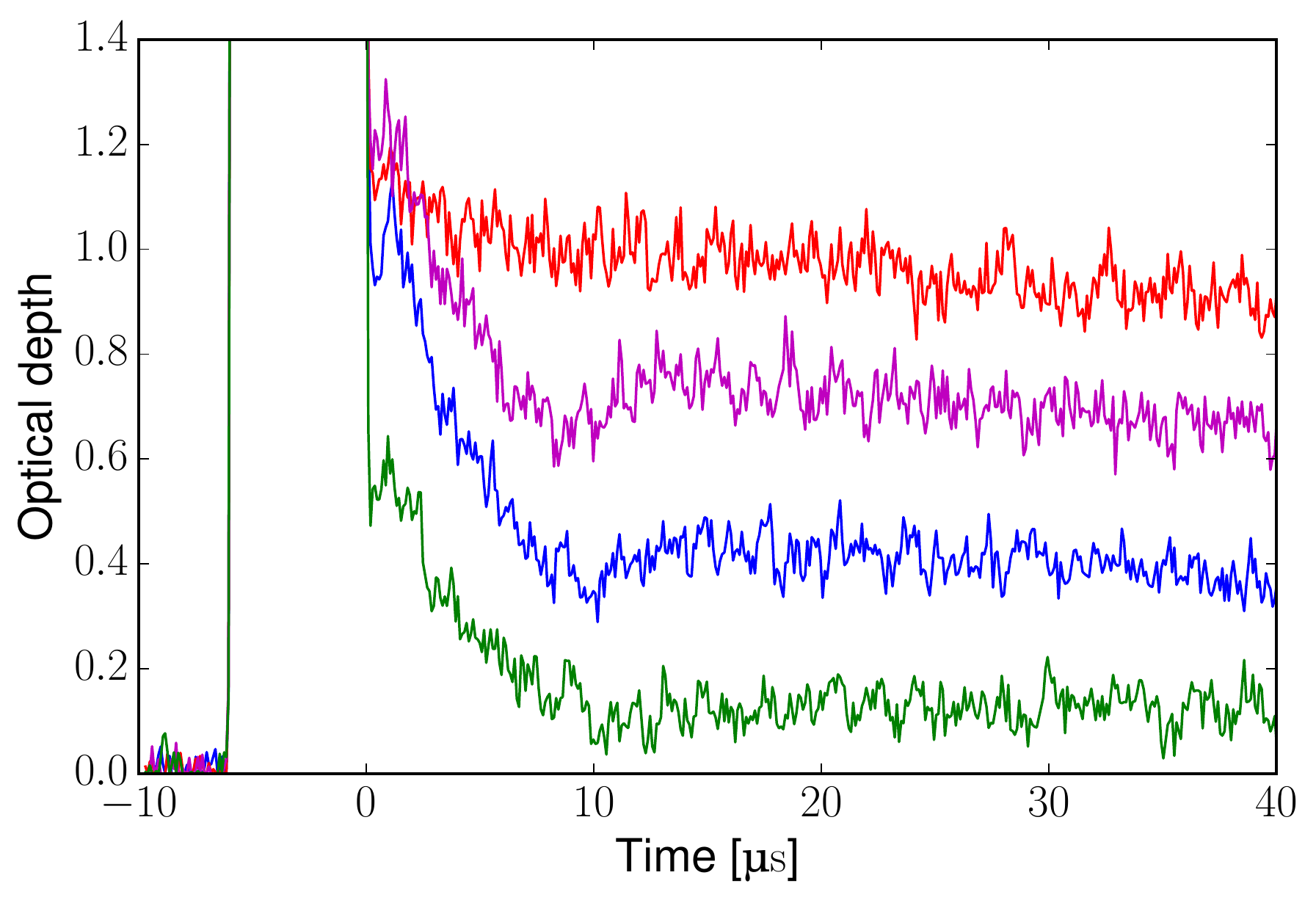}
    \caption{\label{fig:od} Extracted optical depth corresponding to
      the traces shown in \fref{fig:trm}.}
  \end{figure}

  In \fref{fig:od}, we present the optical depth $\alpha$ extracted
  from the corresponding transmittance shown in \fref{fig:trm}.  The
  structured ensemble displays rich transmittance dynamics: During the
  life-time of the Bragg-reflection ($t \approx 0\ldots\SI{1}{\micro
    \second}$) a rise in optical density is observed as the reflection
  decays. This is followed by a decrease of optical depth, which is
  more pronounced for stronger structuring fields. Part of the initial
  rise stems from the decay of axial ordering leading to more
  scattering into free space modes.  Another part of the initial rise
  can also be understood by the continuing inward radial motion of
  atoms increasing their coupling to the TOF guided modes. The
  subsequent decrease stems for fairly weak structuring fields mainly
  from atoms being reflected off the steep repulsive barrier and
  escaping towards the shallow part of the trap potential away from
  the fiber, where they are barely coupled before returning towards
  their start position after one oscillation cycle. Hints of
  oscillatory dynamics with a matching period of
  $t=\SI{12}{\micro\second}$ can be seen in the blue and purple traces
  of \fref{fig:od}.  For stronger structuring fields a significant
  fraction of the atoms gains enough radial kinetic energy to climb
  over the repulsive barrier and collide with the fiber surface, at
  which point atoms cease to interact resonantly with the probe
  light. This leads to slightly faster and more pronounced loss of
  optical depth for strong structuring fields.  In addition to radial
  kinetic energy the atoms also receive axial kinetic energy, which,
  together with the non-separability of the true trap potential, makes
  atoms with enough total energy never return to the vicinity of the
  fiber. The loss dynamics is essentially stationary after a little
  more than one radial oscillation cycle. We note that we observe
  quite similar dynamics in the simulation model, despite the
  simplified treatment of axial and radial motion, see
  \fref{fig:trm_zoom}.  The final ($t>\SI{20}{\micro\second}$) slow
  decrease of optical depth is also observed in the absence of
  structuring light and is due to the limited trap lifetime.
  \begin{figure}[htbp]
    \includegraphics[width=\columnwidth]{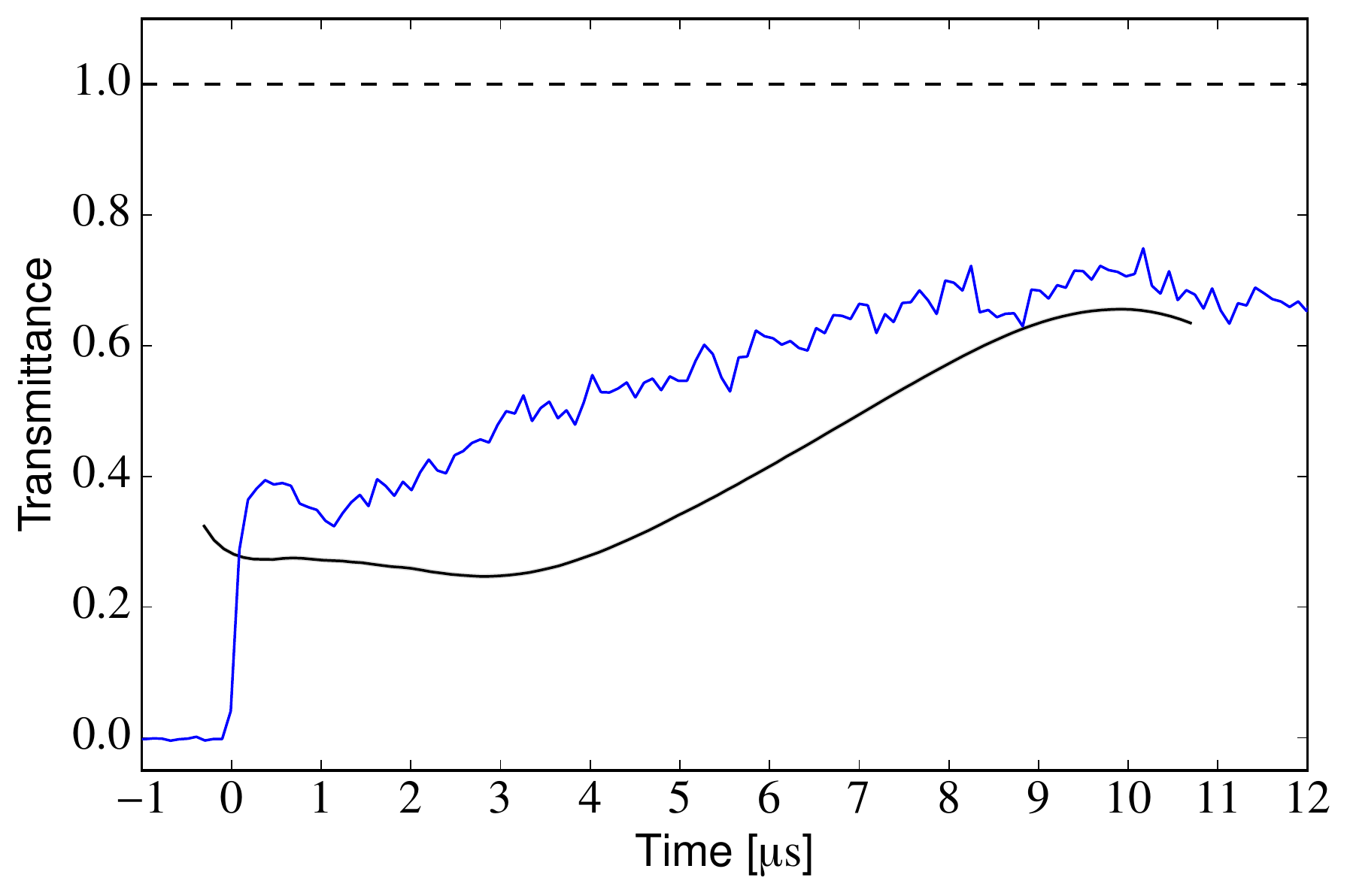}
    \caption{\label{fig:trm_zoom} Zoom of blue transmittance curve in
      \fref{fig:trm} together with the theoretical prediction
      (black).}
  \end{figure}
%

  \section[]{Transfer matrix}
  \label{sec:TransferMatrix}
  For the theoretical modeling of the data presented here and in the
  main text we employ the transfer matrix formalism.
  \begin{figure}[htb]
    \includegraphics[width=.25\columnwidth]{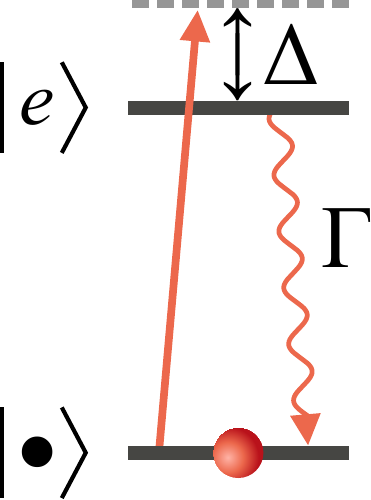}
    \caption{\label{fig:level_diagram} Two-level diagram.}
  \end{figure}
  Each atom is modeled as a two-level system in the low saturation
  approximation characterized by a single scattering parameter
  \begin{gather}
    \beta_j =\frac{\GammaWG} {\Gamma'-2i(\Delta+\Delta_j)},
  \end{gather}
  where the total radiative decay rate is given by
  $\Gamma=\GammaWG+\Gamma'$ with $\GammaWG=\odperatom\Gamma/2$ and
  $\Gamma'$ being the decay rates into the TOF mode and all other
  modes respectively, $\Delta$ is the probe detuning from atomic
  resonance (\fref{fig:level_diagram}), and $\Delta_j$ is an
  additional shift that accounts for the inhomogeneous broadening of
  the atomic transition. Each $\Delta_j$ is drawn from a Gaussian
  distribution with the width $\sigma_\Delta$ (see the following
  section).

  The transfer matrix for the $j$-th atom is
  \begin{gather}
    M_{\text{a},j} =
    \begin{pmatrix}
      1-\beta_j & -\beta_j\\
      \beta_j & 1 + \beta_j
    \end{pmatrix}.
  \end{gather}
  If the position of the atom is $z_j$, then $M_{\text{a},j}$ relates
  $E_\text{R}$ and $E_\text{L}$ (right- and left-going electric fields
  respectively) at the position $z_j^-=z_j-\epsilon$ (just to the left
  of the atom with an infinitesimal $\epsilon$) to the fields at
  $z_j^+=z_j+\epsilon$ (just to the right of the atom). Writing this
  as an equation we have
  \begin{gather}
    \begin{pmatrix}
      \ER(z_j^+)\\
      \EL(z_j^+)
    \end{pmatrix}
    =M_{\text{a},j}
    \begin{pmatrix}
      \ER(z_j^-)\\
      \EL(z_j^-)
    \end{pmatrix}.
  \end{gather}

  The transfer matrices for free propagation between the atoms are
  \begin{gather}
    M_{\text{f},j}=
    \begin{pmatrix}
      e^{ikd_j} & 0\\
      0 & e^{-ikd_j}
    \end{pmatrix},
  \end{gather}
  where $d_j$ is the propagation distance between atom $j$ and atom
  $j+1$, i.e.  $z_{j+1}=z_j+d_j$. For the matrix $M_{\text{f},j}$, it
  holds that
  \begin{gather}
    \begin{pmatrix}
      \ER(z_{j+1}^-)\\
      \EL(z_{j+1}^-)
    \end{pmatrix}
    =M_{\text{f},j}
    \begin{pmatrix}
      \ER(z_j^+)\\
      \EL(z_j^+)
    \end{pmatrix}.
  \end{gather}

  In the model, we neglect the effects of the axial trap potential and
  take the atoms to be initially uniformly distributed along the
  length of the ensemble. The procedure to obtain $d_j$'s is as
  follows:
  \begin{itemize}
  \item $N$ random numbers $\{x_1, x_2, \ldots, x_N\}$ are drawn from
    the uniform distribution on the interval $[0,1)$.
  \item The positions of the atoms are given by $z_j=Lx_j$, where $L$
    is the total length of the ensemble.
  \item The distances between the atoms are then given by
    $d_j=z_{j+1}-z_{j}$, where we define $z_{N+1}=L$.
  \end{itemize}

  In addition, each atom is assigned a random axial thermal velocity
  drawn from a Gaussian distribution according to the temperature of
  the ensemble. Likewise, initial random radial positions and
  velocities are chosen.

  To model the effect of hyperfine pumping, each atom is assigned a
  probability $p_j$ to remain in the $\ket{\bullet}$ state after the
  structuring pulse, with pumping strength quantified by the
  dimensionless parameter $\zeta$, given by
  \begin{equation}
    p_j = \e^{-\zeta\cos^2(2\pi z_j/\ldp^{\text{TOF}})}\,,
    \label{eq:probability}
  \end{equation}
  for the case of atoms not saturated by the structuring light. The
  probability for being pumped from $\ket{\bullet}$ to $\ket{\circ}$
  is thus $1-p_j$. In the numerical implementation, this is taken into
  account by drawing $N$ random numbers from a uniform distribution on
  $[0,1)$ and comparing them with $p_j$ for each atom. The probability
  of such a random number to be smaller than $p_j$ is equal to $p_j$
  and the probability to be bigger than $p_j$ is equal to $1-p_j$. The
  advantage of this approach is that once the array of the random
  numbers is fixed, the depumping process becomes deterministic in
  this model for any choice of the pumping strength $\zeta$. For the
  depumped atoms, $M_{\text{a},j}$ is replaced by the identity matrix.
  We remark that for the short duration and range of intensities of
  the structuring pulses chosen in the experiment the amount of
  hyperfine pumping is quite limited, e.g. at pulse parameters for
  highest reflectance less than \SI{10}{\percent} of the atoms are
  pumped into the $\ket{\circ}$ level. It is, nevertheless, important
  to take the effect into account for good quantitative agreement
  between model and experiment.

  The effect of dipole forces in axial and radial directions is
  modelled by evaluating for each atom the received momentum kick
  according to position in the structuring pulse standing wave with
  the given duration, detuning and intensity.  Isotropic distribution
  over Zeeman levels is assumed and all excited hyperfine levels of
  the atomic transition are taken into account. Position and velocity
  at the instant of probing is then determined applying the
  propagators for ballistic (harmonic) motion in the axial (radial)
  direction.  Each atom the receives updated $\GammaWG$ and $\Gamma'$
  values for the current radial distance to the fiber according to the
  shape of the evanescent guided modes.  For an atom $j$, which has
  come closer to the fiber than a cut-off radius of
  $\SI{10}{\nano\meter}$ at any time between structuring and the
  instant of probing, the transfer matrix $M_{\text{a},j}$ is replaced
  by the identity matrix.

  The theoretical curves in the main text are obtained by averaging
  over typically 100 realizations in the following way:
  \begin{itemize}
  \item For each realization the transfer matrix for the ensemble is
    found
    \begin{gather}
      M_\text{ensemble}=M_{\text{f},N}M_{\text{a},N}\ldots
      M_{\text{f},1}M_{\text{a},1}.
    \end{gather}
  \item If $M_\text{ensemble}$ is written out as
    \begin{gather}
      M_\text{ensemble} =
      \begin{pmatrix}
        M_{11} & M_{12}\\
        M_{21} & M_{22}
      \end{pmatrix},
    \end{gather}
    then the (amplitude) reflection coefficient for the whole ensemble
    is given by $r=M_{12}/M_{22}$. The (amplitude) transmission
    coefficient is $t=1/M_{22}$.
  \item As the final step, we take the mean of the absolute square of
    the reflection (transmission) coefficient $|r|^2$ ($|t|^2$) to
    obtain the reflectance (transmittance).
  \end{itemize}

  Averaging over many ensemble realizations still gives stochastic
  results, but the standard deviation of the mean reflection and
  transmission coefficients can be decreased by using a larger number
  of ensemble realizations and hence can be made arbitrarily small.

  \section[]{Inhomogeneous broadening}
  \label{sec:Inhomogeneous}
  \begin{figure}[tb]
    \includegraphics[width=\columnwidth]{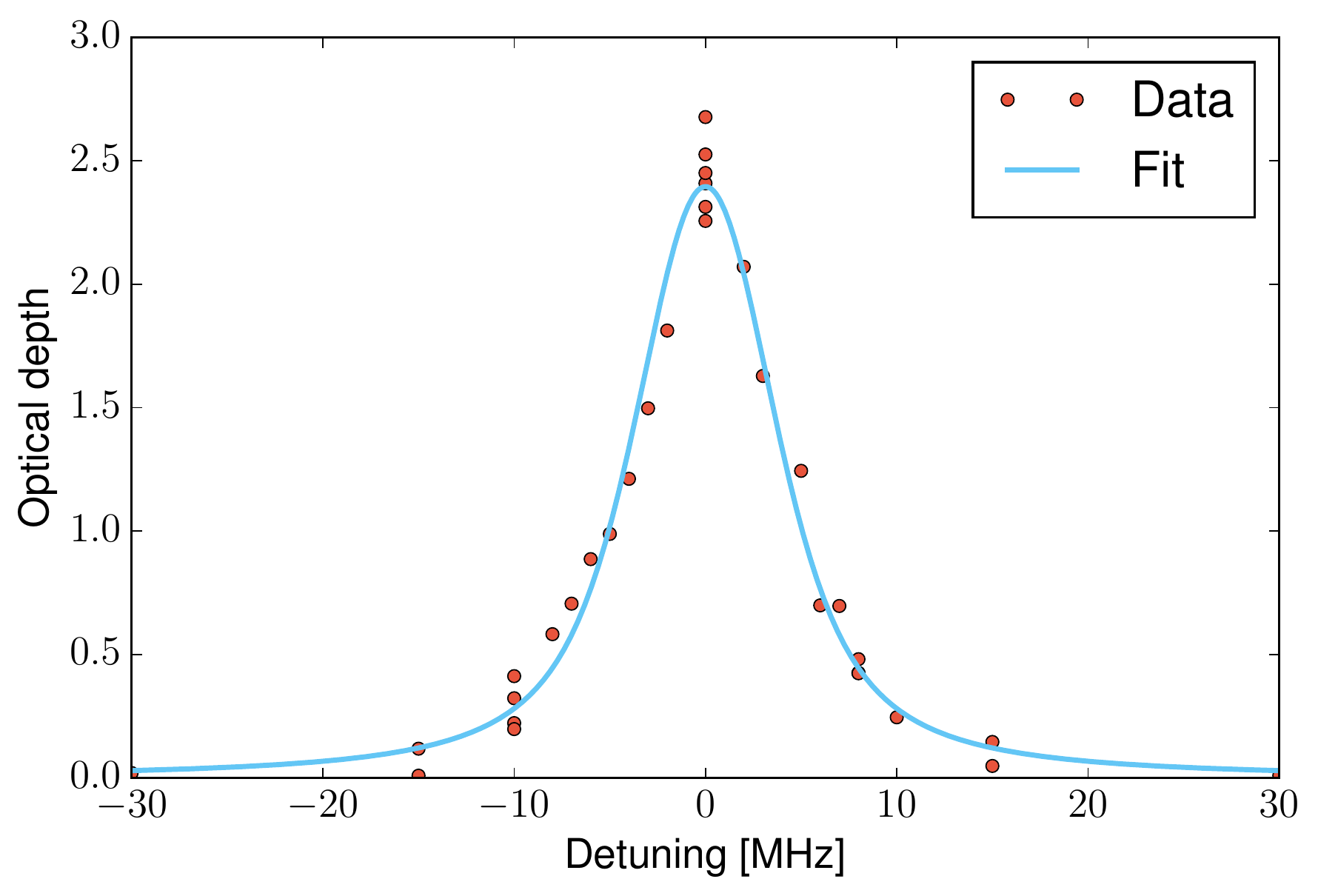}
    \caption{\label{fig:sDelta_fit} Optical depth as a function of the
      probe detuning. Data points are derived from the transmission
      signals for the reflectance data presented in
      \fref[b]{fig:R_vs_detuning_and_fraction} in the main text.}
  \end{figure}
  To estimate the amount of inhomogeneous broadening of the atomic
  transition, we fit the inferred optical depth as a function of the
  probe detuning (\fref{fig:sDelta_fit}) with a Voigt profile:
  \begin{align}
    V(\Delta; \sigma_\Delta, \alpha) &= \int_{-\infty}^{\infty}
    G(\Delta ', \sigma_{\Delta}) L(\Delta-\Delta ') \ud \Delta'\,.
  \end{align}
  Here $G$ and $L$ are the Gauss- and Lorentz-distributions given by:
  \begin{align}
    G(\Delta ', \sigma_{\Delta}) &=
    \frac{1}{\sqrt{2\pi}\sigma_{\Delta}}
    \exp{\bigg(-\frac{\Delta^2}{2\sigma_{\Delta}^2}} \bigg)\,,\\
    L(\Delta-\Delta ') &= \frac{\alpha}{1 +
      (\Delta-\Delta')^2/(\Gamma/2)^2 + s }\,. %
  \end{align}
  $s=P/\Psat=\SI{150}{pW}/\SI{750}{pW}=0.2$ is the saturation
  parameter. The free parameters in the fit are the homogeneous
  on-resonant optical depth $\alpha$, and the inhomogeneous broadening
  $\sDelta$.
}
\end{document}